\documentclass[twocolumn]{aastex62}
\usepackage{amsmath,amstext}
\usepackage{apjfonts} 
\usepackage{chngcntr}
\usepackage{verbatim}

\newcommand{\be}{\begin{eqnarray}}
\newcommand{\ee}{\end{eqnarray}}
\newcommand{\lp}{\left(}
\newcommand{\rp}{\right)}
\newcommand{\lb}{\left[}
\newcommand{\rb}{\right]}

\begin{document}

\normalsize


\title{\Large \textbf{Inferring the Presence of Tides in Detached White Dwarf Binaries}}

\author{Anthony L. Piro}

\affil{The Observatories of the Carnegie Institution for Science, 813 Santa Barbara St., Pasadena, CA 91101, USA; piro@carnegiescience.edu}

\begin{abstract}
Tidal interactions can play an important role as compact white dwarf (WD) binaries are driven together by gravitational waves (GWs). This will modify the strain evolution measured by future space-based GW detectors and impact the potential outcome of the mergers. Surveys now and in the near future will generate an unprecedented population of detached WD binaries to constrain tidal interactions. Motivated by this, I summarize the deviations between a binary evolving under the influence of only GW emission and a binary that is also experiencing some degree of tidal locking. I present analytic relations for the first and second derivative of the orbital period and  braking index. Measurements of these quantities will allow the inference of tidal interactions, even when the masses of the component WDs are not well constrained. Finally, I discuss tidal heating and how it can provide complimentary information.
\end{abstract}


\section{Introduction}

The merger of compact white dwarf (WD) binaries has been hypothesized to generate an incredible range of astrophysical systems and/or events. This includes sdB/O stars \citep{Saio00,Saio02}, R Cor Bor stars \citep{Webbink84}, fast radio bursts \citep{Kashiyama13}, magnetic and DQ WDs \citep{Garcia12}, millisecond pulsars \citep{Bhattacharya91}, magnetars \citep{Usov92}, a subset of gamma-ray bursts \citep{Dar92,Metzger08}, Type Ia supernovae \citep{Iben84}, a source of ultra-high energy cosmic rays \citep{Piro16}, and AM CVn binaries \citep{Postnov06}. An important uncertainty in connecting specific binaries to these outcomes is the role of tidal interactions \citep{Marsh04}\footnote{Although see the arguments by \citet{Shen15} that the vast majority of these binaries may merge.}. Furthermore, these binaries are among the strongest gravitational wave (GW) sources in our Galaxy and will be prime targets for future space-based GW detectors \citep[e.g.,][]{Nelemans09,Marsh11,Nissanke12,Tauris18}. Tides will again be key in determining the orbital evolution detected by these observations.

Many double WD binaries are transferring mass, which makes it difficult to isolate the impact of tidal interactions. Detached WD binaries provide the perfect laboratory for measuring these effects. The sample has grown with surveys such as ELM \citep{Kilic12}, SPY \citep{Napiwotzki04}, and ZTF \citep{Bellm19}, and will only accelerate in size over the next decade with SDSS-V \citep{Kollmeier17} and LSST \citep{LSST,Korol17}. In particular, there is now a binary with a period of $12.75\,{\rm min}$, SDSS J065133.338+284423.37 \citep[hereafter J0651,][]{Brown11}, and another with a period of $6.91\,{\rm min}$, ZTF J153932.16+502738.8 \citep[hereafter J1539][]{Burdge19}, which should be especially helpful for studying the role of tides.

Motivated by these issues, in the following I provide analytic relations that summarize how a binary evolving due to GW emission changes with the degree of tidal locking. These will simplify the interpretation and isolate the impact of tides in future binary WD observations. This is similar to  discussions in \citet{Piro11}, but here I focus on relations that will help constrain tidal interactions empirically rather than following the time dependent orbital evolution. Some similar relations are also provided in \citet{Shah14}, but these use approximations from \citet{Benacquista11}. Here I present exact relations that can be expressed analytically.

In Section~\ref{sec:first derivative}, I summarize the main equations determining the binary evolution and derive the tidal corrections to the first derivative of the orbital period, while in Section~\ref{sec:second derivative}, I focus on the second derivative. In Section \ref{sec:braking}, I discuss the braking index $n$, defined by $\dot{\Omega}\propto \Omega^n$ where $\Omega$ is the binary orbital frequency. In Section \ref{sec:tidal locking}, I discuss how the tidal locking may change with time, and what impact this has on the system's evolution. Simple estimates for these expressions are presented in Section~\ref{sec:estimates}. In Section \ref{sec:measure}, {I discuss the duration of observations needed to measure deviations} in the inspiral from just GW emission. The role of tidal heating in constraining the degree of tidal locking is described in Section~\ref{sec:heating}, and I conclude in Section~\ref{sec:conclusions} with a summary.

\section{First Derivatives of the Orbital Period}
\label{sec:first derivative}


Consider a binary with orbital separation $a$ composed of WDs with masses $M_1$ and $M_2$ and orbital frequency $\Omega^2 = GM/a^3$ where $M=M_1+M_2$. The orbital angular momentum is $J_{\rm orb} = (Ga/M)^{1/2}M_1M_2$. The angular momentum of the system decreases due to GW emission at a rate
\be
	\dot{J}_{\rm gw} = -\frac{32}{5} \frac{G^3}{c^5} \frac{M_1M_2M}{a^4}J_{\rm orb}.
\ee
Taking the time derivative of $J_{\rm orb}$ and setting it equal to $\dot{J}_{\rm gw}$ results in a period derivative of
\be
	\dot{P}_{\rm gw} = -\frac{96}{5} \frac{G^3}{c^5} \frac{M_1M_2MP}{a^4}
	= \frac{3\dot{J}_{\rm gw}}{J_{\rm orb}}P,
	\label{eq:pdotgw}
\ee
if GW emission is acting alone without any tides.

The total angular momentum of the binary system is
\be
	J_{\rm tot} &=& J_{\rm orb} + J_{\rm wd}
	\nonumber
	\\
	&=& (Ga/M)^{1/2}M_1M_2 + I_1\Omega_1+I_2\Omega_2,
\ee
where $I_i$ and $\Omega_i$ are the moments of inertia and spins of the WDs, respectively. To simply include the impact of tidal interactions, I introduce the variable $\eta$, the tidal locking factor, and set $\Omega_1=\Omega_2=\eta\Omega$. When $\eta=0$, there is no tidal effects, and $\eta=1$ corresponds to being completely tidally locked. In detail, there should be a separate $\eta$ for each WD, since they could have different degrees of tidal locking. Given the current level of measurements, it is simpler to consider a single $\eta$ that represents the whole binary system. The total angular momentum becomes
\be
	J_{\rm tot} = (Ga/M)^{1/2}M_1M_2 + \eta(I_1+I_2)(GM/a^3)^{1/2}.
	\label{eq:jtot}
\ee
Taking the time derivative of this and setting $\dot{J}_{\rm tot}=\dot{J}_{\rm gw}$,
\be
	\dot{J}_{\rm gw}
	= \frac{1}{2}\frac{\dot{a}}{a} J_{\rm orb}
	-\frac{3}{2}\frac{\dot{a}}{a} J_{\rm wd}
	+\frac{\dot{\eta}}{\eta}J_{\rm wd}
	\label{eq:angular momentum}
\ee
Using the relation that $\dot{P}/P = (3/2)\dot{a}/a$, I find that when tides are included the period derivative is
\be
	\dot{P}_{\rm tide} = \frac{\dot{P}_{\rm gw}-3(\dot{\eta}/\eta)(J_{\rm wd}/J_{\rm orb})P}{1-3J_{\rm wd}/J_{\rm orb}}.
	\label{eq:pdottides}
\ee
{Note that I use the subscript ``tide'' to denote when {\it both} tides and GWs are acting on the binary. Also, since $J_{\rm wd}\propto \eta$, the second term in the numerator is well-behaved even for $\eta=0$.}

Equation (\ref{eq:pdottides}) demonstrates how tides cause the period to decrease more rapidly than with GW emission alone because angular momentum is being taken from the WD orbits and put into the individual WDs. If the tidal locking is increasing as the orbital period shrinks ($\dot{\eta}>0$), then the period decreases even more rapidly. Similar relations are presented in \citet{Benacquista11} and \citet{Burkart13} for the case $\eta=1$ with $\dot{\eta}=0$. Note in the former work (which used the variable $\Delta_{\rm I}=J_{\rm wd}/J_{\rm orb}$) their expression for the change in the frequency derivative is only correct to order $J_{\rm wd}/J_{\rm orb}$. Equation~(\ref{eq:pdottides}) ignores  deviations due to the energy it takes to make a tidal bulge, since this effect is small \citep{Benacquista11}.

The angular momentum ratio $J_{\rm wd}/J_{\rm orb}$ continually comes up when considering tidal corrections. Useful ways to express this include
\be
	\frac{J_{\rm wd}}{J_{\rm orb}}
	= \frac{\eta(I_1+I_2)M}{M_1M_2a^2} = \frac{\eta (I_1+I_2)M^{1/3}\Omega^{4/3}}{G^{2/3}M_1M_2},
\ee
which is used for some of the estimates below.


\section{Second Derivatives of the Orbital Period}
\label{sec:second derivative}

I next consider the impact of tidal interactions on the second derivative of the orbital period. For GW emission alone, taking the derivative of Equation (\ref{eq:pdotgw}) results in
\be
	\ddot{P}_{\rm gw} = -4 \frac{\dot{a}}{a} \dot{P}_{\rm gw} + \frac{\dot{P}}{P} \dot{P}_{\rm gw} 
	= -\frac{5}{3}  \frac{\dot{P}}{P} \dot{P}_{\rm gw},
\ee
{where I have been careful to distinguish between $\dot{P}$ (the exact derivative of $P$) and $\dot{P}_{\rm gw}$ (the derivative when just GWs are considered) because this difference will be important for subsequent discussions.} Since in this case $\dot{P} = \dot{P}_{\rm gw}$, then
\be
	\ddot{P}_{\rm gw} = -\frac{5}{3}  \frac{\dot{P}_{\rm gw}^2}{P}.
	\label{eq:doubledot pgw}
\ee
Given that $\ddot{P}$ will be difficult to measure, I focus on the simpler scenario where $\eta=1$ with $\dot{\eta}=0$. To find the second derivative including tides, it is helpful to first rewrite Equation (\ref{eq:pdottides}) as
\be
	\dot{P}J_{\rm orb} -3\dot{P}J_{\rm wd} = \dot{P}_{\rm gw}J_{\rm orb}.
\ee
{Taking the derivative of both sides of this expression,
\be
	\ddot{P}J_{\rm orb} + \frac{1}{2}\frac{\dot{a}}{a}\dot{P}J_{\rm orb}
	- 3\ddot{P} J_{\rm wd} - 3\frac{\dot{\Omega}}{\Omega}\dot{P}J_{\rm wd}
	\nonumber
	\\
	= -\frac{5}{3}  \frac{\dot{P}}{P}\dot{P}_{\rm gw}J_{\rm orb} + \frac{1}{2}\frac{\dot{a}}{a}\dot{P}_{\rm gw}J_{\rm orb}.
\ee
Rewriting all the first derivatives in terms of $\dot{P}$, dividing by $J_{\rm orb}$, and then collecting similar terms, the final result is}
\be
	\ddot{P}_{\rm tide} \left( 1 - \frac{3J_{\rm wd}}{J_{\rm orb}}  \right)
	= -\frac{4}{3}  \frac{\dot{P}_{\rm tide}\dot{P}_{\rm gw}}{P}
	- \frac{\dot{P}_{\rm tide}^2}{P} \left( \frac{1}{3} + \frac{3J_{\rm wd}}{J_{\rm orb}}  \right).
	\nonumber
	\\
	\label{eq:ddotptides}
\ee
When combined with Equation~(\ref{eq:pdottides}) for $\dot{P}_{\rm tide}$ (using $\dot{\eta}=0$ for consistency), this expression provides an analytic relations for $\ddot{P}_{\rm tide}$.
Comparing Equations~(\ref{eq:doubledot pgw}) and (\ref{eq:ddotptides}) shows that tidal interactions make the second derivative even more negative than the GW only case.

\section{Braking Index}
\label{sec:braking}

Beyond the values of the first and second derivative, another way to think about tides is in terms of a braking index. The basic idea is analogous to pulsars \citep{Shapiro83} where one wishes to measure $n$ such that
\be
	\dot{\Omega} \propto \Omega^n.
\ee
The main difference is that here $\Omega$ refers to the orbital frequency of the binary rather than the pulsar spin frequency (and perhaps it should be referred to as an ``acceleration index'' here), although also see discussions of the braking index by \citet{Nelemans04} and \citet{Stroeer05} in the context of AM CVn systems. It is straightforward to show
\be
	n = \frac{\Omega\ddot{\Omega}}{\dot{\Omega}^2}
	= 2 - \frac{P\ddot{P}}{\dot{P}^2}.
\ee
Using the relations from Sections \ref{sec:first derivative} and \ref{sec:second derivative}, for GWs only \citep{Webbink98},
\be
	n_{\rm gw} = 11/3,
	\label{eq:ngw}
\ee
while when tides are included
\be
	n_{\rm tide} = \frac{10}{3}+\frac{1/3 +3J_{\rm wd}/J_{\rm orb}}{1-3J_{\rm wd}/J_{\rm orb}}.
	\label{eq:ntide}
\ee
Thus fitting for the power law $\dot{\Omega}\propto \Omega^n$ in real systems could be another way to infer the presence of tides. Although the exact value of $n$ depends on the specific parameters of the binary and the degree of tidal locking, simply showing that $n>11/3$ would demonstrate tides are occurring without having to know these parameters.

\section{The Rate of Change of Tidal Locking}
\label{sec:tidal locking}

An additional factor in $\dot{P}_{\rm tide}$ that is often not considered is $\dot{\eta}$, the rate of change of the tidal locking. This will depend in detail on the model for the tidal interaction. In \citet{Piro11}, the tide is treated using a parameterization with a standard quality factor $Q$. When this tidal interaction is integrated forward in time, it is found that the binary reaches an equilibrium spin at any given point where the ratio of the tidal forcing frequency to the orbital frequency is roughly the ratio of the tidal synchronization time $\tau_{\rm tide}$ to the GW inspiral time $\tau_{\rm gw} = -(3/2)P/\dot{P}$, i.e.,
\be
	1-\eta \approx \frac{\tau_{\rm tide}}{\tau_{\rm gw}}.
\ee
Furthermore, this work finds that for constant $Q$, $\tau_{\rm tide}/\tau_{\rm gw}\propto P^{1/3}$. This ratio thus gets smaller at shorter orbital period, resulting in a more tidally locked binary. In reality, this will depend on the details of how the tides act, and so I consider a general form of $\tau_{\rm tide}/\tau_{\rm gw}\propto P^{\beta}$ with $\beta>0$.

Assuming a model of this form and taking the derivative results in,
\be
	\dot{\eta} \approx -\beta\frac{\dot{P}}{P} \frac{\tau_{\rm tide}}{\tau_{\rm gw}}
	\approx -\beta\frac{\dot{P}}{P} (1-\eta).
\ee
This particular model has a couple of important consequences for $\dot{P}$. First, it makes $\dot{P}$ even more negative. Second, it adds corrections to $\dot{P}$ even when $\eta\approx0$ because it implies a rapidly changing $\eta$, while for $\eta\approx1$ this model gives $\dot{\eta}\approx0$. In the future, it would be useful to see how large $\dot{\eta}$ might be for more physical models, such as dynamical tides when the WDs are near a resonance \citep[e.g.,][]{Fuller11}.

\section{Estimates for Real Systems}
\label{sec:estimates}

Equations (\ref{eq:pdottides}) and (\ref{eq:ddotptides}) summarize the analytic expressions for the first and second period derivatives with  tidal interactions, and Equation (\ref{eq:ntide}) the braking index. It is helpful to estimate what values are implied by these expressions.

Low mass WD binaries are typical composed of a He and C/O WD. Taking typical values of $M_1=0.25\,M_\odot$ and $M_2=0.5\,M_\odot$, for GW emission alone,
\be
	\dot{P}_{\rm gw} = -1.2\times10^{-11}M_{0.25}M_{0.5}M_{0.75}^{-1/3}P_{10}^{-5/3}{\rm s\,s^{-1}},
\ee
where $M_{0.25}=M_1/0.25\,M_\odot$, $M_{0.5}=M_2/0.5\,M_\odot$, $M_{0.75}=M/0.75\,M_\odot$, and $P_{10}=P/10\,{\rm min}$. The fractional change when tides are included to first order in $J_{\rm wd}/J_{\rm orb}$ when $\dot{\eta}=0$ is
\be
	\frac{\dot{P}_{\rm tide}-\dot{P}_{\rm gw}}{\dot{P}_{\rm gw}}
	\approx \frac{3J_{\rm wd}}{J_{\rm orb}}
	\approx  0.096\eta I_{51}M_{0.25}^{-1}M_{0.5}^{-1}M_{0.75}^{1/3}P_{10}^{-4/3},
	\nonumber
	\\
	\label{eq:pdot_est}
\ee
where $I_{51}=(I_1+I_2)/10^{51}\,{\rm g\,cm^2}$, which is estimated using a moment of inertia of $I_i \approx 0.2 M_iR_i^2$ \citep{Marsh04}. If one instead assumes $\dot{\eta}$ follows the model described in Section \ref{sec:tidal locking},
\be
	\frac{\dot{P}_{\rm tide}-\dot{P}_{\rm gw}}{\dot{P}_{\rm gw}}
	\approx \lb\beta +(1-\beta)\eta \rb \frac{3J_{\rm wd}(\eta=1)}{J_{\rm orb}}.
\ee
Even for $\eta=0$, this model gives a deviation in the period derivative.

The second derivative due to GW emission alone is calculated from Equation (\ref{eq:doubledot pgw}) to be
\be
	\ddot{P}_{\rm gw} = -3.9\times10^{-25}M_{0.25}^2M_{0.5}^2M_{0.75}^{-2/3}P_{10}^{-13/3}\,{\rm s\,s^{-2}}.
\ee
To estimate the second derivative with tides, consider Equation (\ref{eq:ddotptides}) in the limit  $J_{\rm wd}\ll J_{\rm orb}$,
\be
	\ddot{P}_{\rm tide} &\approx& 
	-\frac{4}{3}  \frac{\dot{P}_{\rm tide}\dot{P}_{\rm gw}}{P} \left( 1 + \frac{3J_{\rm wd}}{J_{\rm orb}}  \right)
	\nonumber
	\\
	&&- \frac{\dot{P}_{\rm tide}^2}{P} \left( \frac{1}{3} + \frac{3J_{\rm wd}}{J_{\rm orb}}  \right)\left( 1 + \frac{3J_{\rm wd}}{J_{\rm orb}}  \right).
\ee
Substituting for $\dot{P}_{\rm tide}$ using Equation (\ref{eq:pdottides}), and collecting terms first order in $J_{\rm wd}/J_{\rm orb}$ results in
\be
	\ddot{P}_{\rm tide}
	\approx -\frac{5}{3}  \frac{\dot{P}_{\rm gw}^2}{P}
	\left( 1 + \frac{42}{5}\frac{J_{\rm wd}}{J_{\rm orb}}  \right),
\ee
or a fractional change in the second derivative of the orbital period of
\be
	\frac{\ddot{P}_{\rm tide}-\ddot{P}_{\rm gw}}{\ddot{P}_{\rm gw}}
	\approx \frac{42}{5}\frac{J_{\rm wd}}{J_{\rm orb}} \approx 0.27\eta I_{51}M_{0.25}^{-1}M_{0.5}^{-1}M_{0.75}^{1/3}P_{10}^{-4/3}.
	\nonumber
	\\
	\label{eq:pddot_est}
\ee
Thus the deviation in the second derivative should be more pronounced (by a factor of a few) than the first derivative.

The braking index estimated to first order in  $J_{\rm wd}/J_{\rm orb}$ is
\be
	n_{\rm tide} &\approx &\frac{11}{3}+ \frac{4J_{\rm wd}}{J_{\rm orb}}
	\nonumber
	\\
	&\approx& 11/3+ 0.13 \eta I_{51}M_{0.25}^{-1}M_{0.5}^{-1}M_{0.75}^{1/3}P_{10}^{-4/3}.
\ee
Changes in $n$ are at a level of around ten percent. Unlike comparing $\dot{P}_{\rm gw}$ and $\dot{P}_{\rm tide}$, $n_{\rm gw}$ is independent of the masses of the WDs, so any deviation from $11/3$ would be strong evidence for tides. This expression is $\sim50\%$ larger than the estimate in \citet{Shah14}.

\section{Measuring the Braking Index}
\label{sec:measure}

The previous sections show that measuring $n$ (or equivalently $\ddot{P}$) provides a way to infer the presence of tides without having to know the WD masses. It is natural to ask how long it will take to measure $n$ with sufficient accuracy. Taylor expanding the orbital phase of the binary (assuming $\phi=0$ at $t=0$)
\be
	\phi = \Omega t + \frac{1}{2}\dot{\Omega} t^2 +\frac{1}{6}\ddot{\Omega} t^3 + \cdots.
\ee
If the uncertainty in the eclipse timing of a binary is $\delta t$, then the uncertainty in the phase is $\delta\phi = \Omega \delta t$. The fractional uncertainty in constraining the frequency derivative can be read off from the second term of the Taylor expansion to be
\be
	\frac{\delta \dot{\Omega}}{\dot{\Omega}} \approx \frac{2\delta \phi}{\dot{\Omega}t^2},
\ee
where $t$ the length of time of the observation, while the uncertainty in the braking index is
\be
	\frac{\delta n}{n} \approx \frac{\delta\ddot{\Omega}}{\ddot{\Omega}} \approx \frac{6\delta\phi}{\ddot{\Omega}t^3}.
\ee
Using the $\ddot{\Omega}$ for GW emission alone, this results in
\be
	\frac{\delta n}{n} \approx 1.3 M_{0.25}^{-2}M_{0.5}^{-2}M_{0.75}^{2/3}P_{10}^{16/3} \delta t_{10}\lp \frac{t}{10\,{\rm yrs}}\rp^{-3},
	\label{eq:delta n}
\ee
where $\delta t_{10}=\delta t/10\,{\rm ms}$ for the eclipse timing accuracy. The strong scalings with $P$ and $t$ mean that $n$ will be much easier to measure for short orbital period systems that are observed for a long time. In particular, for J1539 a $\approx10\%$ constraint on $n$ should be possible after $\approx12\delta t_{10}^{1/3}\,{\rm yrs}$.

\section{Constraining Tides with Tidal Heating}
\label{sec:heating}

Although direct measurements of the first and second period derivative are the cleanest way to infer tidal interactions, such measurements require observations over a long time baseline. It is therefore useful to have other complementary methods such as tidal heating.



The total energy of the WD binary is composed of orbital and spin components,
\be
	E_{\rm tot} = E_{\rm orb} + E_{\rm wd} = -\frac{GM_1M_2}{2a} + \frac{1}{2}(I_1+I_2)\eta^2\Omega^2.
\ee
Taking the time derivative of this,
\be
	-\frac{\dot{a}}{a}E_{\rm orb} + 2\frac{\dot{\Omega}}{\Omega}E_{\rm wd} +2\frac{\dot{\eta}}{\eta}E_{\rm wd} 
	= \dot{E}_{\rm gw} -L_{\rm tide},
	\label{eq:energy}
\ee
where $L_{\rm tide}$ is energy lost to tidally heating the WDs (defined to be positive here) and
\be
	\dot{E}_{\rm gw} = -\frac{32}{5}\frac{G^4}{c^5}\frac{M_1^2M_2^2M}{a^5},
\ee
is the energy lost to GWs. Combining Equations (\ref{eq:angular momentum}) and (\ref{eq:energy}), results in
\be
	L_{\rm tide} &=& 2\lp \frac{\dot{\eta}}{\eta}-\frac{\dot{P}}{P}\rp\left( E_{\rm wd} + E_{\rm orb}\frac{J_{\rm wd}}{J_{\rm orb}}\right)
	\nonumber
	\\
	&= &\lp \frac{\dot{\eta}}{\eta}-\frac{\dot{P}}{P}\rp (I_1+I_2) \Omega^2\eta(1-\eta),
	\label{eq:ltide1}
\ee
for the tidal heating rate. A rough estimate can be made from just considering $\dot{P}$ from {GWs \citep{Iben98},}
\be
	L_{\rm tide} &\approx& -\frac{\dot{P}_{\rm gw}}{P}(I_1+I_2)\Omega^2
	\nonumber
	\\
	&\approx& 2.2\times10^{33}I_{51}M_{0.25}M_{0.5}M_{0.75}^{-1/3}P_{10}^{-14/3}
	{\rm erg\,s^{-1}},
	\label{eq:ltide2}
\ee
but the exact amount depends on the degree of tidal locking and the model for how the tidal locking changes with time.

\begin{figure}
\epsscale{1.2}
\plotone{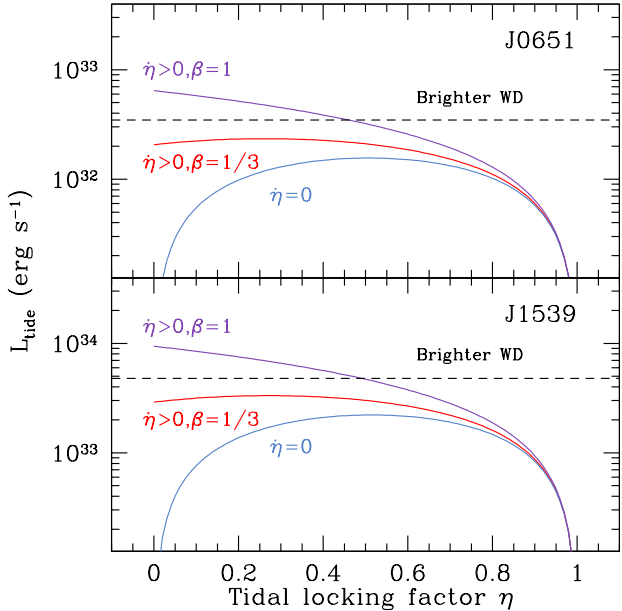}
\caption{The tidal heating rate given by Equation (\ref{eq:ltide1}) for the cases of  J0651 \citep{Hermes12} and J1539 \citep{Burdge19} in the upper and lower panels, respectively. In each case, the blue curves assume $\dot{\eta}=0$, while the red and purple curves use the model from Section \ref{sec:tidal locking} for $\dot{\eta}$ with $\beta=1/3$ and $\beta=1$, respectively. The dashed horizontal lines show the currently observed luminosities of the brighter  component of each binary, while the dimmer component would be below the lowest plotted luminosities.}
\label{fig:ltide_eta}
\epsscale{1.0}
\end{figure}

In the case of $\dot{\eta}=0$, the heating rate scales as
\be
	L_{\rm tide}\propto \eta(1-\eta),
	\label{eq:ltide no eta dot}
\ee
Thus, in the limit of extremely weak tidal locking ($\eta\approx 0$) the energy losses go all into GWs and for strong tidal locking (\mbox{$\eta\approx1$}) the energy losses go into spinning up the WDs, and in each case the tidal heating is small \citep[also see the discussion in][]{Fuller12}. For $\eta=1/2$, the heating is maximum. If the WD spins are allowed to vary independently (so that $\Omega_1=\eta_1\Omega$ and $\Omega_2=\eta_2\Omega$), then the heating simply scales with $I_1\eta_1(1-\eta_1)+I_2\eta_2(1-\eta_2)$ instead, and heating is maximum for $\eta_1=\eta_2=1/2$. 

If one uses the model for $\dot{\eta}$ from Section \ref{sec:tidal locking}, the heating scales instead scales as
\be
	L_{\rm tide}\propto [\beta+(1-\beta)\eta](1-\eta).
\ee
This again goes to zero for $\eta=1$ because all of the extra energy goes into spinning up the WDs, but now at $\eta=0$ there is a non-zero amount of heating because the tidal locking factor is changing rapidly at small $\eta$. In addition, $L_{\rm tide}$ will be larger for the $\dot{\eta}\neq0$ case because $\dot{P}$ is more negative as well.


Figure \ref{fig:ltide_eta} shows how the tidal heating changes with $\eta$ using Equation (\ref{eq:ltide1}). I consider cases where $\dot{\eta}=0$ (blue curves) and when $\dot{\eta}\neq 0$ with either $\beta=1/3$ (red curves) or $\beta=1$ (purple curves). The $\beta=1/3$ case corresponds to a constant $Q$ model \citep{Piro11}, while $\beta=1$ is chosen because it results in $L_{\rm tide}\propto 1-\eta$, similar to the most extreme tidal heating expected \citep{Fuller12}. The upper and lower panels use parameters (masses, radii, and orbital periods) appropriate for J0651 and J1539, respectively. In each case, the brighter WD could be consistent with tidal heating, but it depends on the details of the tidal dissipation. If $\dot{\eta}=0$ or if $\beta$ is too small, then the tidal heating is insufficient to explain the observed luminosities. The similarity of the top and bottom panels potentially indicates that similar physical processes are occurring to dissipate the tides in each of the systems. It is curious that in this picture the C/O WD rather than the He WD would be more tidally heated in J1539, but perhaps this can be accomplished via resonances with a dynamical tide. 

Alternatively, the large luminosities of the brighter WDs may simply mean that the observed emission cannot be dominated by tides. Instead, the relative luminosities of these WDs could reflect their {age \citep{Istrate14,Istrate16}}. It may seem somewhat paradoxical that the C/O WD would be younger in J1539, but there are binary scenarios where this can occur \citep{Toonen12}. These binaries would then have to be generated with periods fairly close what is observed now, perhaps explaining how surveys were able to find them so close to merger (one would expect many more long period WD binaries if they start these longer periods, although there may be selection effects that favor short periods). Another possibility discussed by \citet{Burdge19} is that the C/O WD has undergone recent accretion.

For both J0651 and J1539, the dimmer WDs argue that there is little tidal heating (they are both so dim to be below the luminosities plotted in Figure \ref{fig:ltide_eta}). Given the symmetrical dependence on $\eta$ in Equation (\ref{eq:ltide no eta dot}), they would have to either have $\eta\lesssim 0.005$ or $\eta\gtrsim 0.995$. An important caveat is that this assumes the tidal heating can be readily radiated by the WDs. This is not unreasonable given that the WD thermal time is typically shorter than the GW merger time and that dynamical tides are expected to be dissipated closer to the surface \citep[rather than deep in the WD interior,][]{Fuller11,Fuller12,Fuller13}, so that the energy can be radiated more easily. Given their extremely low luminosities though, it may mean that the tidal heating is getting trapped within the WDs.


\section{Conclusions and Discussion}
\label{sec:conclusions}

In this work, I have presented analytic relations for the impact of tidal interactions on $\dot{P}$, $\ddot{P}$, the braking index, and heating for binaries driven together by GWs. The $\dot{P}$ measured for J0651 and J1539 cannot currently answer whether tidal interactions are occurring because the masses of the WDs are not known to sufficient accuracy. The braking index $n$, defined by $\dot{\Omega}\propto \Omega^n$, is $11/3$ for GWs only and larger by an amount summarized by Equation~(\ref{eq:ntide}) when tides are {also} involved. Measuring $n\neq11/3$ would be a way to infer the presence of tides without requiring precisely measured WD masses.  Assessing how well our ability to constrain $n$ improves with time in Equation (\ref{eq:delta n}) shows that it will take on the order of $\approx 10\,{\rm yrs}$ to make such measurements for the shortest known detached eclipsing binaries. These simple prescriptions are useful for assessing the ability of space-based GW detectors to infer the presence of tides \citep[e.g.,][]{Shah14,Littenberg19}.

Measuring tidal heating is another way to constrain tidal interactions. The brighter WDs in both J0651 and J1539 have luminosities that could be explained by tidal heating, but this depends on the tidal dissipation model used. This makes it difficult to directly constrain the influence of tides from the luminosities alone. The dimmer WDs must either be nearly tidally locked or not locked at all under the assumption that their tidal heating can be readily radiated.  Having a wider sample of detached WDs with different orbital periods in the future will hopefully help determine how well WD luminosities can be used to constrain tidal interactions.

\acknowledgments
I thank Jim Fuller, J.~J.~Hermes, Thomas Kupfer, Thomas Marsh, Thomas Tauris, and Silvia Toonen for helpful discussions and feedback. I also thank the organizers of The Beginning and Ends of Double White Dwarfs meeting in Copenhagen, Denmark (July 1-5, 2019), where some of this work was inspired.

\bibliographystyle{apj}

\end{document}